\documentclass[sigconf,nonacm,screen]{acmart}

\usepackage{subcaption}
\usepackage{tikz}
\AtBeginDocument{%
  }

\begin{document}

\title{Investigating the Effect of Prior Exposure and Fidelity on Quality and Realism Perception of VR Digital Twins}

\author{Maximilian Warsinke}
\email{warsinke@tu-berlin.de}
\orcid{0009-0004-0264-5619}
\affiliation{
  \institution{Quality and Usability Lab\\ Technische Universität Berlin}
  \city{Berlin}
  \country{Germany}
}

\author{Maurizio Vergari}
\email{maurizio.vergari@tu-berlin.de}
\orcid{0000-0002-2640-6979}
\affiliation{%
  \institution{Quality and Usability Lab\\ Technische Universität Berlin}
  \city{Berlin}
  \country{Germany}
}

\author{Tanja Kojić}
\email{tanja.kojic@tu-berlin.de}
\orcid{0000-0002-8603-8979}
\affiliation{%
  \institution{Quality and Usability Lab\\ Technische Universität Berlin}
  \city{Berlin}
  \country{Germany}
}

\author{Daniel Nikulin}
\email{nikulin@tu-berlin.de}
\orcid{0009-0002-7985-5441}
\affiliation{%
  \institution{Quality and Usability Lab\\ Technische Universität Berlin}
  \city{Berlin}
  \country{Germany}
}

\author{Sebastian Möller}
\email{sebastian.moeller@tu-berlin.de}
\orcid{0000-0003-3057-0760}
\affiliation{%
  \institution{Quality and Usability Lab\\ Technische Universität Berlin, DFKI}
  \city{Berlin}
  \country{Germany}
}


\begin{abstract}
This study explores how prior exposure to physical objects influences the quality and realism perception of Digital Twins (DT) with varying levels of fidelity in Virtual Reality (VR). In a mixed experimental design, 24 participants were divided into two equal groups: an exposure group, in which members were shown physical objects before inspecting and rating their replicas in VR, and a control group without prior knowledge. Three objects were presented, each under four fidelity conditions with varying texture resolution and geometric detail. Participants rated perceived quality and realism through in-VR self-reports. Statistical analysis revealed that texture resolution significantly affected realism and quality perception, whereas geometric detail only influenced quality ratings. Investigating the between-factor, no significant effect of exposure on quality and realism perception was found. These findings raise important questions about the cognitive relationship between physical objects and their digital counterparts and how fidelity influences the perception of DTs in VR.
\end{abstract}



\newcommand\copyrighttext{%
  \footnotesize \textcopyright\ 2025 ACM. This is the author's version of the work. It is posted here for your personal use. Not for redistribution. The definitive Version of Record will be published in Proceedings of the 31st ACM Symposium on Virtual Reality Software and Technology (VRST ’25). The published article is available at: \href{https://dl.acm.org/doi/10.1145/3756884.3766019}{https://dl.acm.org/doi/10.1145/3756884.3766019}.%
}

\newcommand\copyrightnotice{%
\begin{tikzpicture}[remember picture,overlay,shift={(current page.south)}]
  \node[anchor=south,yshift=10pt] at (0,0) {\fbox{\parbox{\dimexpr\textwidth-\fboxsep-\fboxrule\relax}{\copyrighttext}}};
\end{tikzpicture}%
}





\maketitle

\copyrightnotice

\section{Introduction}

\begin{figure}
    \centering
    \includegraphics[width=0.9\linewidth]{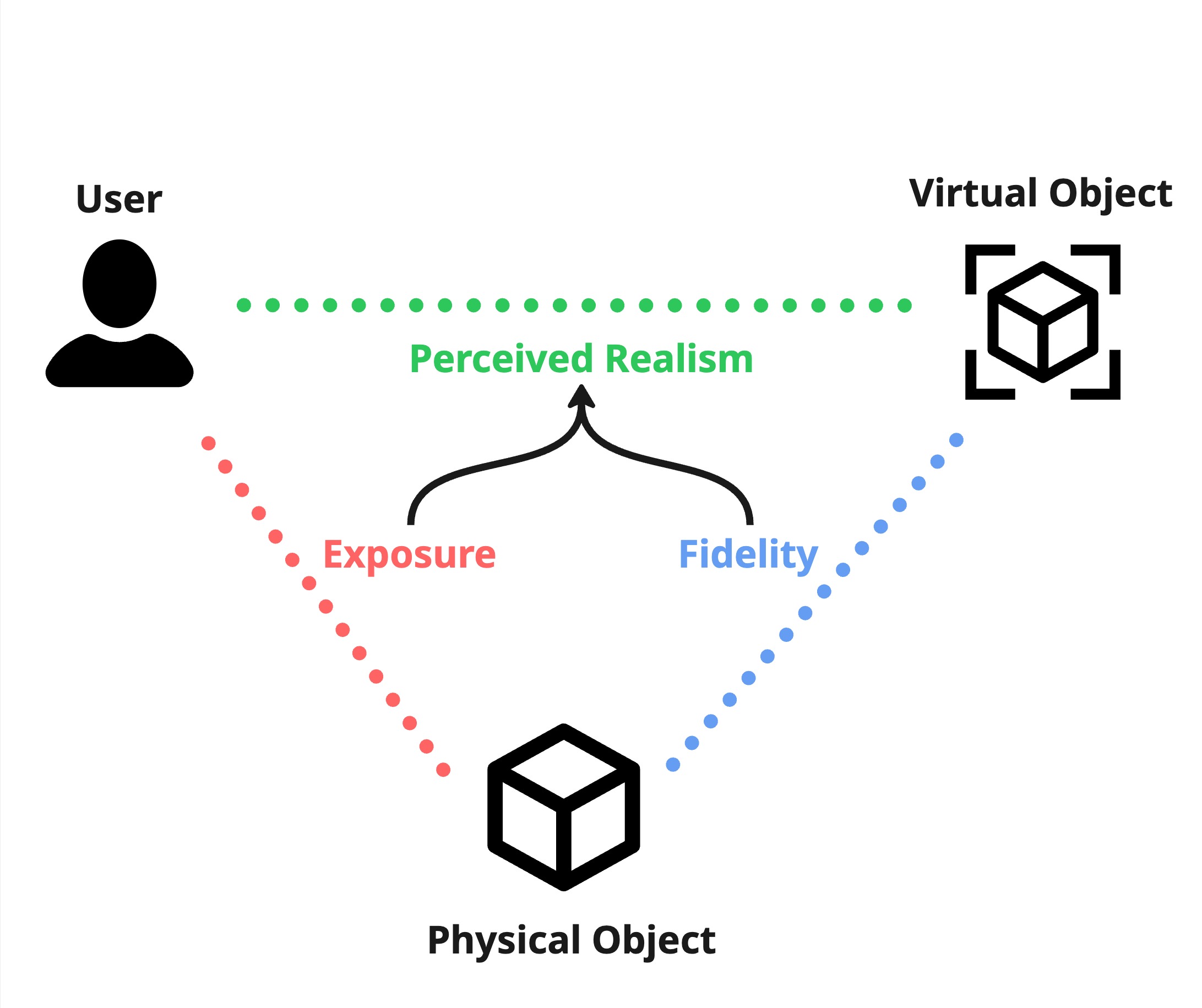}
    \caption{Hypothesized model of the relationship between exposure, fidelity, and perceived realism of DTs in VR.}
    \label{fig:scema}
\end{figure}

Virtual Reality (VR) is increasingly used as a medium to interact with Digital Twins (DTs) because of its immersive capabilities. A DT is a replica of an object, location, or system, characterized by being linked to a physical counterpart through a data stream \cite{Jones2020}. In recent years, DTs have been employed in nearly all fields imaginable, including manufacturing, healthcare, and education \cite{singh2022}.

While DTs are predominantly visualized in 3D space, they can vary in their \textit{fidelity}—"the objective degree of exactness with which real-world experiences and effects are reproduced by a computing system" \cite{mcmahan2012, gerathewohl1969fidelity}. The ultimate goal of VR is to provide high-fidelity experiences \cite{al2022framework}, aiming to enhance \textit{perceived realism}—"the user’s individual judgment about the degree of realism of the VE" \cite{Weber2021} (VE refers to Virtual Environment). \textit{Quality} judgments involve the user's comparison of perceived quality features to desired quality features \cite{perkisQUALINETWhitePaper2020a}.



In applications centered around visuals, such as architecture \cite{narasimha2019empirical} or VR tourism \cite{beck2019virtual}, the aim is to create the highest possible degree of realism perception, which can be approximated with higher-fidelity reconstructions. In VR training applications, users can learn how to operate industrial machines on DTs \cite{Rukangu2021} or practice evacuation scenarios in a safe environment \cite{Loh2023}. Here, the focus lies on how users interact with the VR environments and the resulting learning outcomes. Interestingly, research has shown that high visual realism has a positive impact on user experience \cite{Goncalves2023} and that user responses are more realistic \cite{Slater2009a}. In manufacturing or smart city contexts, the objective is clear information and pragmatic visuals rather than high fidelity \cite{zhu2019}. For example, visualization methods, such as color-coding based on energy consumption \cite{ruohomaki2018}.

Creating large-scale, high-fidelity DTs can be costly. Manual modeling requires substantial labor from skilled 3D artists, whereas photogrammetry approaches rely on advanced hardware, computational power, and data collection \cite{Pentangelo2024}. On the user side, realism perception is shaped by multiple factors, such as material, texture, lighting, and consistency \cite{SchmiedKowarzik2024}. This raises the question of the optimal fidelity that minimizes replication efforts while ensuring a high degree of perceived realism. Therefore, identifying key fidelity predictors for quality and realism perception would improve workflows and inform optimization efforts.

An additional factor influencing quality and realism perception is the user's expectation towards the VR experience. For example, younger adults have been found to report lower perceived realism, possibly because of their familiarity with virtual environments (e.g., through video games) \cite{Dilanchian2021}. Because of a DT's unique property of having a counterpart in the physical world, knowledge about this counterpart might additionally influence the users’ expectations. For instance, it has been shown that knowledge of the size of an object influences the size and depth perception of a virtual replica \cite{rzepka2023}. We assume that this object knowledge also impacts quality and realism perception. Specifically, we argue that users would perceive a virtual object as less realistic when they are familiar with the reference due to heightened expectations (knowledge about the texture and geometric details). This would imply that when developing DTs, additional resources should be allocated to objects and environments that are commonly known by users compared to unusual or imaginative content.

Although studies explored different degrees of fidelity and scene properties and their effect on user experience \cite{Hvass2017, Gutiérrez2020, Brade2021}, it is widely unknown how knowledge about physical counterparts comes into the equation. How users perceive replicated spaces has only recently become a research objective \cite{skarbez022}. A long-term vision, outlined in an early Metaverse roadmap that is still being built upon, describes a "Mirrored World", a perfectly replicated virtual world, as one possible part of a Metaverse \cite{smart2007, yu2022, raveendran2024}. The feeling of users being in such spaces was also defined as "replicated world illusion" \cite{skarbez2021}. These developments create the need to investigate the perceptual characteristics that emerge from the integration of DTs in VR. Specifically, there is a gap in the literature on how this physical and virtual connection impacts the user experience.

A mixed-design experiment was conducted to explore the two dynamics: the impact of different fidelity levels and the effect of object knowledge, induced through prior exposure, on realism and quality perception (cf. Figure \ref{fig:scema}). To approach the topic, it was decided to start on a small scale with a fundamental form of a DT, a replica of a physical object. Three DTs were created using photogrammetry and manually altered for a set of fidelity conditions. Each virtual object was presented in a VR scene with a combination of two levels of geometric fidelity and two levels of texture fidelity. Participants were randomly assigned to either the exposure or control group. The physical reference objects were shown to the exposure group prior to the VR experience, while the control group was not. In the VR application, participants had to inspect and rate their perception of quality and realism.

\begin{figure}
    \centering
    \begin{subfigure}{0.45\textwidth}
        \centering
        \includegraphics[width=\linewidth]{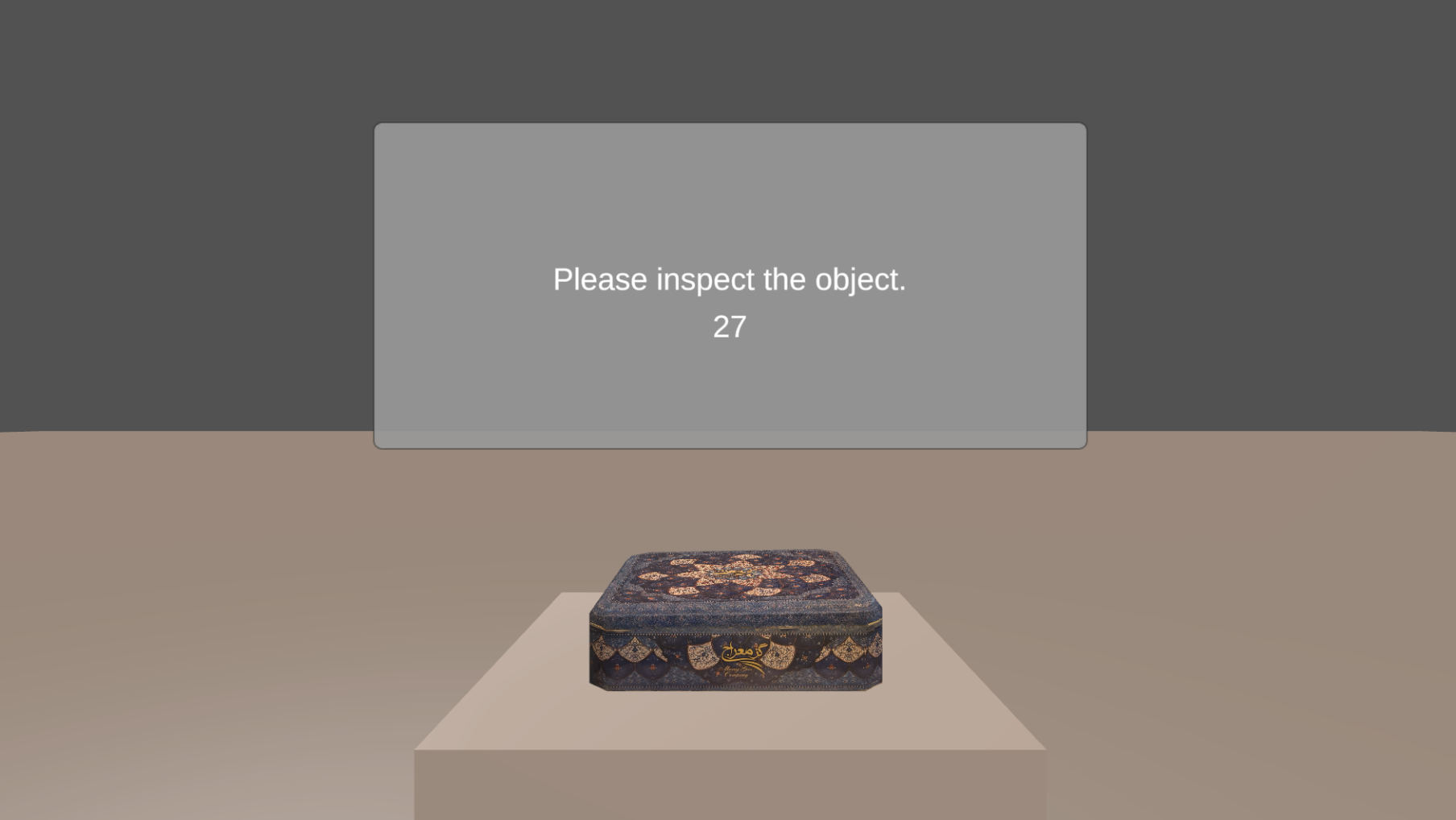}
        \label{fig:notused1}
    \end{subfigure}
    \begin{subfigure}{0.45\textwidth}
        \centering
        \includegraphics[width=\linewidth]{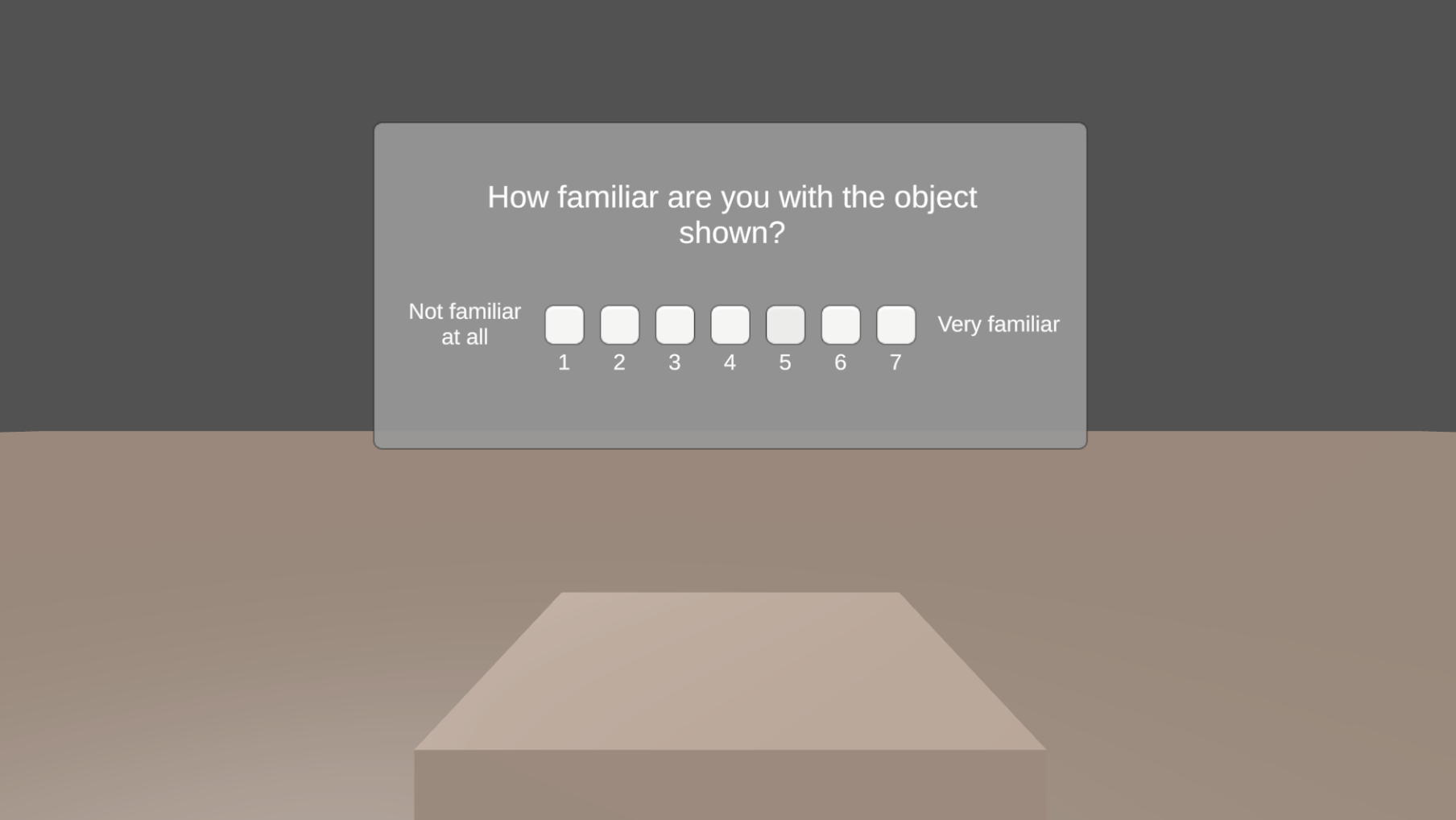}
        \label{fig:notused2}
    \end{subfigure}
    \caption{Screenshots from the VR application showing the inspection phase (top) and the rating phase (bottom).}
    \label{fig:unity}
\end{figure}

\section{Related Work}
\paragraph{\textbf{Concepts}}
In computer graphics, fidelity, visual realism, and photorealism are regarded as similar concepts \cite{Fan2018}. According to Ferwerda, an image is photorealistic if it produces the same visual response as looking at a scene in the analog world \cite{Ferwerda2003}. In the context of simulation, the term fidelity goes back to Gerathewohl, who defined it as "the degree to which a device accurately produces a specific effect". \cite{gerathewohl1969fidelity}. In Mixed Reality (MR), Milgram and Kishino defined the term Reproduction Fidelity as "the quality with which the synthesizing display is able to reproduce the actual or intended images of the objects being displayed" \cite{Milgram1994}. McMahan later made a distinction between display fidelity (reproduction of real-world sensory stimuli) and interaction fidelity (reproduction of real-world interactions) \cite{mcmahan2012}. In a recent publication, Bonfert et al. present a comprehensive taxonomy differentiating between eight types of fidelity \cite{bonfert2025}. The term visual realism is often used in a similar context. Slater et al. distinguish visual realism between geometric realism (geometric fidelity) and illumination realism (lightning fidelity) \cite{Slater2009a}. Visual realism, however, is regarded as an objective descriptor, similar to fidelity, in contrast to perceived realism, which is regarded as the subjective judgment of realism \cite{Weber2021}. Another related subjective concept is the concept of presence in virtual or simulated environments. Articulated by Slater as a combination of place illusion (the sense of being there) and plausibility illusion (the sense that what is happening is real) \cite{Slater2009b}. Despite a conceptual relationship, higher visual realism does not necessarily lead to a greater sense of presence \cite{JungLindeman2021}. However, perceived realism was proposed as one part of a two-dimensional construct of presence together with "being there" \cite{Weber2021}. Finally, Gonçalves et al. proposed a new theoretical model using the term subjective realism, which they clearly distinguish from objective realism \cite{goncalves2025}. In their model, they also draw a connection to presence. Quality or experienced quality in the context of immersive media is defined as "a result of a comparison and judgment process, in which the perceived quality features, resulting from a perception and reflection process triggered by a physical signal, are compared to the desired quality features underlying the user’s expectations"\cite{perkisQUALINETWhitePaper2020a}. Visual realism and quality are interconnected, and Fan et al. referred to the quality evaluation of computer graphics as part of predicting visual realism \cite{Fan2018}. In this research, we treat realism and quality as subjective factors, referring to them as perceived realism and perceived quality throughout, while the objective properties of the virtual world are described in terms of fidelity. The distinction between realism and quality perception is that high-quality visuals do not necessarily have to be realistic (e.g., stylized graphics, fantastic worlds).

\paragraph{\textbf{Measurements}}
To measure perceived realism, researchers mostly rely on subjective measurements through self-reports, sometimes supported by complementary physiological measurements \cite{Goncalves2021}. Schmied-Kowartzki et al. proposed the Visual Realism Classification Scale (VRCS), a comprehensive scale for measuring the visual realism of 3D models in MR, distinguishing between the factors of Lighting, Reflection, Texture, Structure, Form, and Internal Consistency \cite{SchmiedKowarzik2024}. Most commonly, perceived realism is assessed as part of the Igroup Presence Questionnaire (IPQ), a standardized questionnaire for measuring presence, as it is one of its sub-scales \cite{ipq, Goncalves2021}. Quality ratings have separate methodologies from the field of computer vision \cite{vanhoey2017}. Nehmé et al. compared three subjective methods from the field of image processing and applied them to the quality assessment of 3D graphics in VR: Category Rating with Hidden Reference (ACR-HR), Double Stimulus Impairment Scale (DSIS), and Subjective Assessment Methodology for Video Quality (SAMVIQ) \cite{Nehmé2021}. Their findings suggest that DSIS, in which participants are exposed to two 3D models side by side and asked to rate the quality degradation, yielded the most promising results. As fidelity is objective, it can only be measured through objective assessment. One approach is an add-on for the 3D software Blender, developed to generate a level of detail (LOD) score for 3D scenes that distinguishes between geometric and texture fidelity \cite{rodriguezgarcia2025}.

\paragraph{\textbf{User Studies}}
While there is a lack of research on the relationship between fidelity and strictly perceived realism, several studies have explored the relationship between 3D object fidelity and other user experience metrics, predominantly presence. Hvass et al. conducted a study using a video game as a stimulus and tested it with varying degrees of geometric and texture fidelity \cite{Hvass2017}. The results of self-reports and physiological measurements showed that participants rated a higher sense of presence and fear responses in scenes with higher visual realism. Gutiérrez et al. assessed the quality metrics for 3D models under different lightning conditions in MR \cite{Gutiérrez2020}. The models were presented with varying levels of geometry and texture and rated through an overall quality estimation. The findings showed that brighter light is a factor for increased quality ratings. Brade et al. explored how solid colored "CAD-style" objects would compare to fully texturized models in a VR assembly task \cite{Brade2021}. While high-fidelity machines created a significantly higher sense of presence, plain models scored sufficiently well to be considered as an option for lower development costs. Braga et al. explored how the visual fidelity of virtual food items influences users’ perceived realism, task motivation, engagement, and interest \cite{braga2025}. They reported that visual fidelity itself did not affect the assessed metrics. However, a relationship was found between perceived realism, task motivation, engagement and interest. Braga et al. argued that, instead of visual fidelity, the typicality of the experience and plausibility may play prominent roles in shaping the user's perception. Skarbez et al. were among the first to investigate how users perceive virtual replicas of physical spaces \cite{skarbez022}. They conducted three studies manipulating the scale of the replicated room and furniture, removing elements from the scene, and altering lighting conditions. Their findings indicate that the scale of the room and furniture significantly influenced perceived realism, whereas lighting had a minimal impact. Lastly, a study by Rzepka et al. explored the effect of object knowledge, specifically the size of well-known objects (e.g., Rubik's Cube), on VR visual perception \cite{rzepka2023}. Participants were shown multiple virtual objects in varying sizes and distances via a VR headset. Their findings indicate that participants tended to rely on familiar object sizes despite what was displayed in VR. Despite the variety of studies, no previous work has explicitly tested the present metrics. However, the findings highlight that user perception is shaped by a combination of objective properties of the VR and cognitive or contextual factors.

\section{Methods}

\subsection{Hypotheses}
The following hypotheses were formulated regarding the effects of prior exposure and fidelity on perceived quality and realism:

\textbf{Prior Exposure}
\begin{itemize}  
    \item $H_{1}:$ Prior exposure to the physical counterpart will lead to lower perceived quality.
    \item $H_{2}:$ Prior exposure to the physical counterpart will lead to lower perceived realism.
\end{itemize}

\textbf{Texture Resolution}
\begin{itemize}  
    \item $H_{3}:$ Higher texture resolution will lead to a higher perceived quality.
    \item $H_{4}:$ Higher texture resolution will lead to a higher perceived realism.
\end{itemize}  

\textbf{Geometric Detail}
\begin{itemize}  
    \item $H_{5}:$ Higher geometric detail will lead to a higher perceived quality.
    \item $H_{6}:$ Higher geometric detail will lead to a higher perceived realism.
\end{itemize}

\textbf{Correlations}
\begin{itemize}  
    \item $H_{7}:$ Perceived quality and perceived realism are positively correlated.
    \item $H_{8}:$ Participant age and perceived realism are positively correlated.
\end{itemize}

\subsection{Study Design}
The study follows a mixed-design approach (2 $\times$ 3 $\times$ 2 $\times$ 2), testing for effects between independent groups (2 levels) and within-subject effects across three factors (3, 2, and 2 levels), resulting in 12 repeated measures per participant.

\textbf{Between-Subjects Factor}
\begin{itemize}
    \item Prior Exposure: Exposure vs. Control
\end{itemize}

\textbf{Within-Subjects Factors}
\begin{itemize}
    \item Object Type: Box vs. Charger vs. Labeler
    \item Texture Resolution: 
    \begin{itemize}
        \item Low Texture (LT): 256 $\times$ 256 pixels
        \item High Texture (HT): 2084 $\times$ 2084 pixels
    \end{itemize}
    \item Geometric Detail:
    \begin{itemize}
        \item Low Geometry (LG): $\approx$ 100 quads
        \item High Geometry (HG): $\approx$ 40.000 quads
    \end{itemize}
\end{itemize}

A balanced Latin Square was used to minimize order effects, creating a minimum requirement of 12 participants for each group. Participants were randomly assigned to one of two groups. An a priori power analysis was conducted using G*Power \cite{faul2007} to determine the minimum detectable effect sizes for a total sample of 24 participants, with an alpha level of 0.05 and power set at 0.95. The analysis indicated sufficient power to detect within-subjects and interaction effects of at least $f = 0.22$ (partial $\eta^2 \approx 0.046$). For the between-subjects main effect, the design was powered to detect large effects of $f = 0.58$ (partial $\eta^2 \approx 0.252$). The experiment was approved by the Ethics Committee of Technische Universität Berlin.

\subsection{Participants}
24 participants took part in the study (12 per group). The control group consisted of eight men and four women, whereas the exposure group included nine women and three men. The mean age of the control group was 36.08 years (SD = 12.88), and the mean age of the exposure group was similar at 36.92 years (SD = 10.73). The average experience with MR was 3.58 (SD = 1.04) for the control group and 3.25 (SD = 1.30) for the exposure group. The average gaming experience was 2.83 (SD = 1.68) for the control group and 1.91 (SD = 1.45) for the exposure group, and the average experience with 3D modeling software was 2.50 (SD = 1.19) for the control group and 2.17 (SD = 1.47) for the exposure group. All values regarding experience were assessed on a scale of 1 to 7, with 7 being the highest.

\subsection{Technology and Tools}
The stimuli were created through photogrammetric scanning using the free version of the mobile application, Polycam. The captured models were then downloaded and edited using Blender and Instant Meshes \cite{wenzel2015} to create quad-based, game engine-ready models. The low-poly versions of the objects were modeled manually following the shapes of the high-poly reference, as recommended by Pentangelo et al. \cite{Pentangelo2024}. A virtual experimental environment was set up in Unity, and an embedded virtual rating interface was developed. The XR interaction toolkit was used for basic VR functionalities, and the universal render pipeline was used for mobile rendering. The generated mipmap was disabled in the texture settings to avoid dynamic-resolution scaling. The anti-aliasing was set to 8x, the shadows were disabled, and the smoothness of the materials was set to zero. For the experiment, participants used a Meta Quest 3.

\begin{figure}
    \centering
    \includegraphics[width=0.45\textwidth]{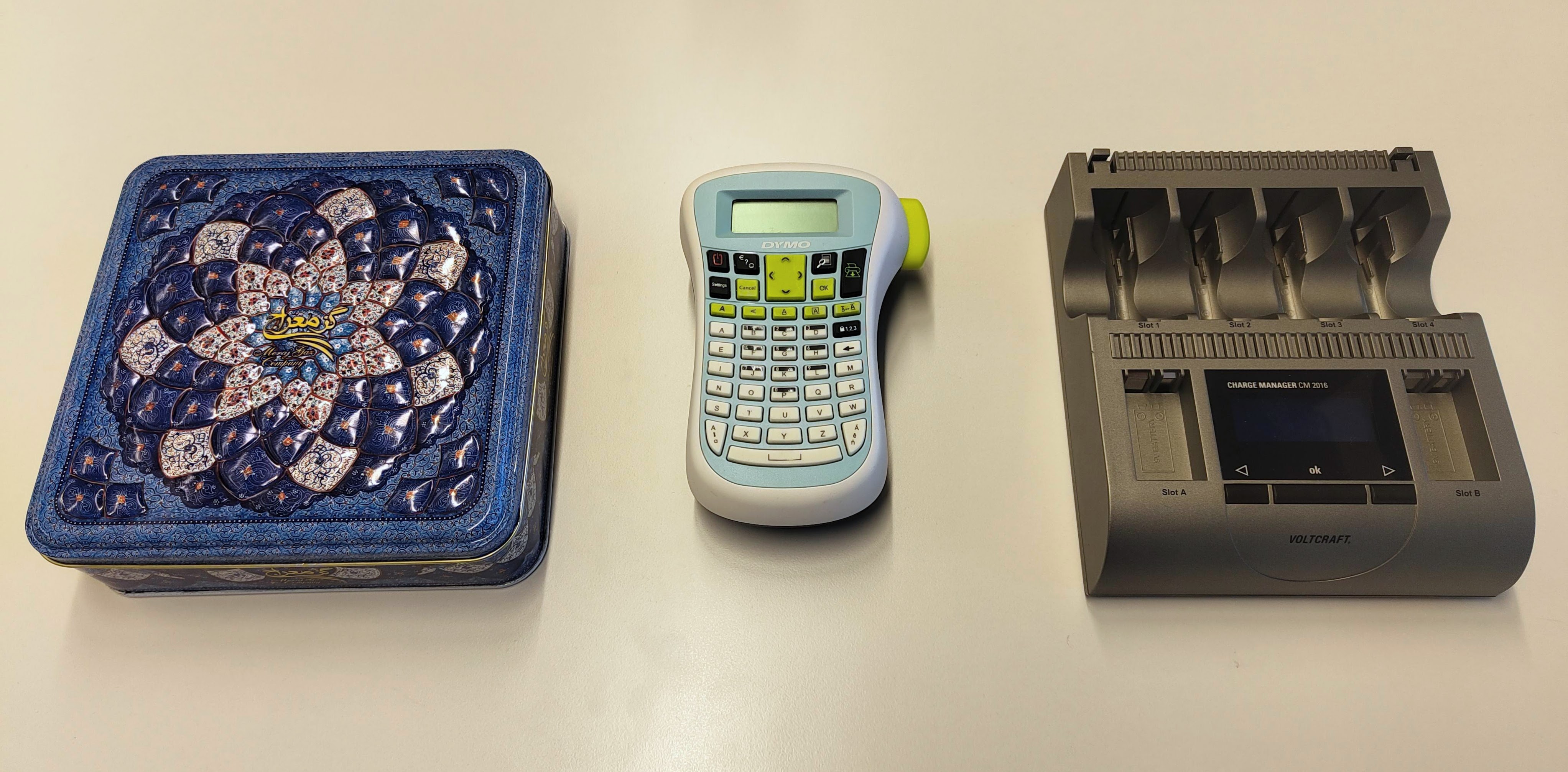}
    \caption{Photograph of the physical objects as shown to the exposure group: Sweets box, labeler, and charger.}
    \label{fig:objects}
\end{figure}

\subsection{Stimuli}
\begin{figure*}
    \centering
    \begin{subfigure}{0.24\textwidth}
        \centering
        \includegraphics[width=\linewidth]{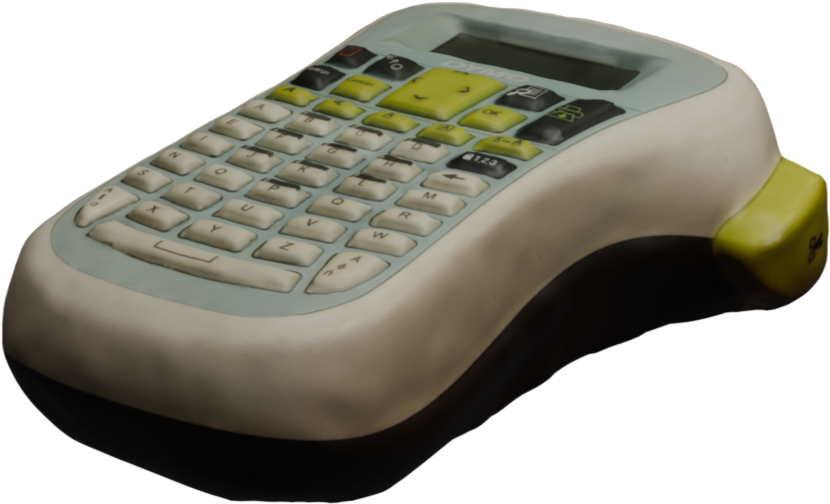}
        \caption{HGHT}
        \label{fig:lab_HGHT}
    \end{subfigure}
    \begin{subfigure}{0.24\textwidth}
        \centering
        \includegraphics[width=\linewidth]{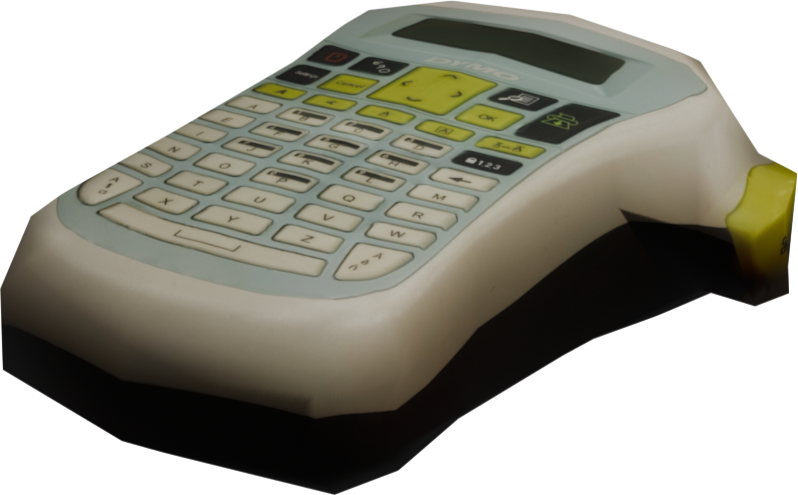}
        \caption{LGHT}
        \label{fig:lab_LGHT}
    \end{subfigure}
    \begin{subfigure}{0.24\textwidth}
        \centering
        \includegraphics[width=\linewidth]{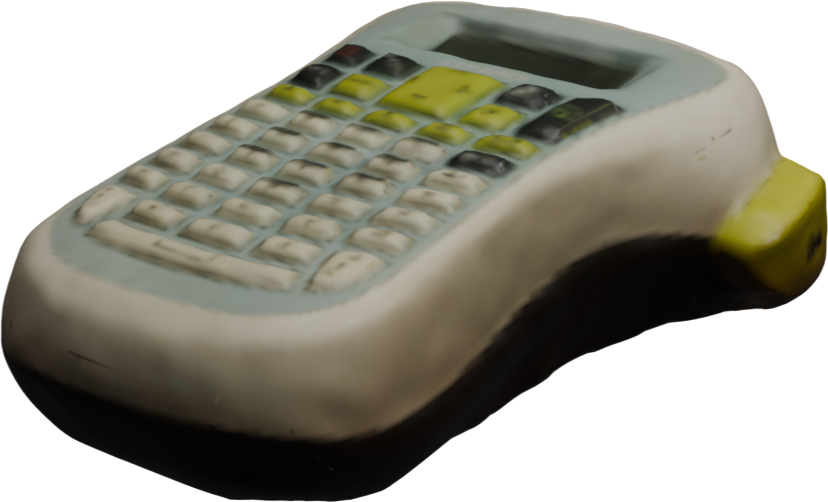}
        \caption{HGLT}
        \label{fig:lab_HGLT}
    \end{subfigure}
    \begin{subfigure}{0.24\textwidth}
        \centering
        \includegraphics[width=\linewidth]{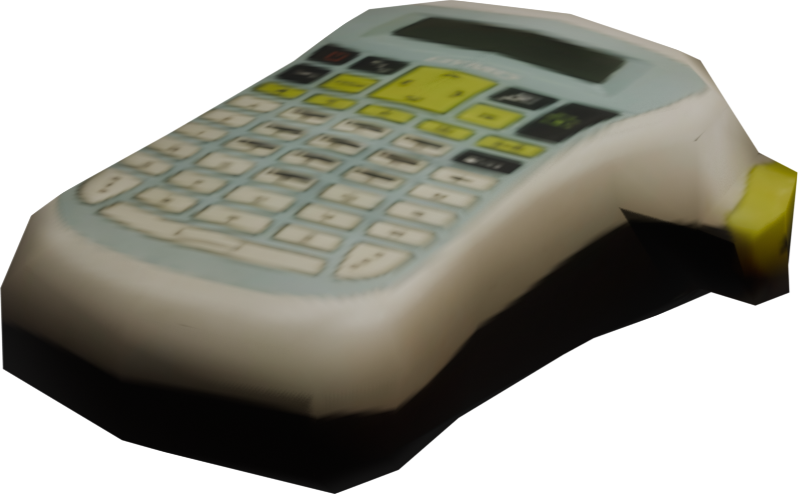}
        \caption{LGLT}
        \label{fig:lab_LGLT}
    \end{subfigure}
    \caption{Renders of the fidelity conditions of the labeler object created with Blender.}
    \label{fig:all_figures}
\end{figure*}

The selected objects were a box of Persian sweets, a labeler for printing custom labels, and a battery charger (see Figure \ref{fig:objects}). Although these objects belong to a conventional office setting, they are not commonly used in everyday tasks, aiming to ensure an equal baseline of familiarity. For the fidelity conditions, this study focused on the geometric detail and texture resolution, as they are pivotal in the creation of DTs. Therefore, shadows and material properties were disabled in Unity to better control the independent variables. The geometric fidelity of the "high" conditions was close to that of the initial 3D scans, with approximately 40000 quads. In this condition, details such as the buttons of the labeler and the convex pattern shapes on the box were noticeable (see Figure \ref{fig:lab_HGHT} and \ref{fig:lab_HGLT}). The geometry of the "low" conditions was highly reduced while still resembling the object in overall shape, with approximately 100 quads. Round shapes of the labeler or the round edges of the box were visibly angular (see Figure \ref{fig:lab_LGHT} and \ref{fig:lab_LGLT}). The "high" texture condition had a resolution of 2084 $\times$ 2084 pixels, meeting the highest possible resolution a Meta Quest 3 can display. Under these conditions, text elements on objects, such as the brand name, were readable and sharp (see Figure \ref{fig:lab_LGHT} and \ref{fig:lab_HGHT}). A resolution of 256 $\times$ 256 pixels was chosen for the "low" condition to create a distinct contrast. In this condition, the overall surfaces appeared blurry, and the text elements were unreadable (see Figure \ref{fig:lab_LGLT} and \ref{fig:lab_HGLT}).

\begin{table*}
    \centering
    {\small
    \begin{tabular}{|c|c|p{7cm}|c|c|}
        \hline
        \textbf{\#} & \textbf{ID} & \textbf{Question} & \textbf{Left Anchor} & \textbf{Right Anchor} \\
        \hline
        Q1 & FAM & How familiar are you with the object shown? & Not familiar at all & Very familiar \\
        \hline
        Q2 & MOS & How would you rate the virtual object's overall quality? & Very low & Very high \\
        \hline
        Q3 & LOD & How would you rate the virtual object's level of detail? & Very low & Very high \\
        \hline
        Q4 & TEX & How would you rate the quality of the virtual object's texture? & Very low & Very high \\
        \hline
        Q5 & REAL1 & How real did the virtual object seem to you? & Not real at all & Completely real \\
        \hline
        Q6 & REAL2 & How much did your experience with the virtual object seem consistent with your real-world experience? & Not consistent & Very consistent \\
        \hline
        Q7 & REAL3 & How real did the virtual object seem to you? & About as real as an imagined object & Indistinguishable from a real object \\
        \hline
        Q8 & REAL4 & The virtual object seemed more realistic than a real object. & Fully disagree & Fully agree \\
        \hline
    \end{tabular}
    }
    \caption{List of the in-VR post-condition questionnaire. All answers were given on a 7-point scale.}
    \label{tab:virtual_object_survey}
\end{table*}

\subsection{Measurements}
A demographic questionnaire assessed age, profession, and gender as well as prior experience with 3D experiences, categorized into three types: experience with mixed reality, experience with gaming, and experience with 3D modeling software. During the experiment, the questions were integrated into the VR application and had to be answered after each condition (See Table \ref{tab:virtual_object_survey}). Q1 aimed to control familiarity with the objects to ensure an equal baseline. Similar to the Mean Opinion Score (MOS), which is widely used in the field of QoE \cite{ITU-P8002-2016}, Q2 aims to assess overall quality. To further distinguish the quality ratings, Q3 and Q4 were introduced to target the fidelity levels of texture and geometry. Q5--Q8 are adaptations of the iGroup Presence Questionnaire (IPQ) \cite{ipq}. While this standardized questionnaire primarily assesses presence, one of the four dimensions deals with perceived realism. This is a common approach because, at present, no standardized questionnaire for measuring realism perception exists \cite{Goncalves2021}. The wording is originally tailored to entire VR scenes and was adapted to match the experiment. The formulations regarding "Virtual World" were replaced with "Virtual Object". The reasoning behind this change was to ensure that participants based their ratings solely on the objects rather than including the VR environment in their judgments. All questions used a 7-point scale with changing anchors.

\subsection{Procedure}
For the exposure group, the three physical reference objects were set up on the table next to the consent forms (similar to Figure \ref{fig:objects}). During the explanation of the experimental procedure, the instructor explained that the objects were 3D-scanned and that they would shortly inspect them in VR. The control group received a neutral explanation, with the objects hidden in a closet and without specification of the types of objects. Participants were then asked to complete the demographic questionnaire. Before starting the application, they were familiarized with the head-mounted device. Typically, in 3D quality rating experiments, short durations for inspection are chosen (6 - 15 seconds \cite{Nehmé2021, guo2017, nehme2021}). However, to allow participants to circumnavigate and inspect the object from all sides, the exposure time was extended to 30 seconds. During pre-testing, this duration seemed to be appropriate. Figure \ref{fig:unity} (top) depicts the virtual scene during the observation phase. After the timer ran out, the object was hidden to ensure an equal inspection time. This also implies that the participants had to rate from their memory. Figure \ref{fig:unity} (bottom) shows the rating user interface. Before the actual questions, the participants had to answer a multiple-choice question regarding one detail of the object. This was the pretended purpose of the inspection and aimed to secure the participant's attention. The entire experiment took approximately 20--30 minutes and was compensated with 10 Euros.

\subsection{Data Analysis}
\label{section:data}
By taking the average of Q5--Q8, a perceived realism score was calculated for each participant, following the scoring approach of the IPQ. The analysis of Q3 and Q4, which aimed to distinguish between texture and geometry quality perception, did not provide additional value, possibly because of unsuitable wording (e.g., "Level of detail" may not be exclusively associated with geometry). They were excluded from the data. For familiarity (Q1), only the participants’ first observation of an object type was of interest for the baseline, and subsequent ratings were therefore excluded.

For statistical testing of between and within-effects on quality and realism perception, each a mixed-design analysis of variance (ANOVA) with a 2 $\times$ 3 $\times$ 2 $\times$ 2 factorial structure was carried out. The assumptions of sphericity and homogeneity were met.
A mixed-design ANOVA was conducted using SPSS for both perceived quality and perceived realism. While Q2 ratings are ordinal, they were treated as approximately interval because they were measured on a 7-point scale. Due to the violation of normality in a few data groups, an additional robust, resampling-based analysis was performed using the R package MANOVA.RM \cite{rm.anova}. In the results section, Wald-Type Statistics (WTS) and ANOVA-Type Statistics (ATS) are therefore reported in addition to the traditional mixed ANOVA results presented in Table \ref{tab:result_quality} and Table \ref{tab:result_realism}.

To investigate relationships between perceived quality, perceived realism and demographic data, a correlation analysis was conducted across the total sample to increase the statistical power. Spearman's rank test was chosen for correlation analysis because the assumption of normality was not met for all measurements. Only the statistically significant variable pairs are presented in the results section. For all analyses, a significance threshold of $\alpha$ = .05 (equivalent to $p$ < .05) was used to determine statistical significance.

\section{Results}

\begin{figure}
    \centering
    \includegraphics[width=0.9\linewidth]{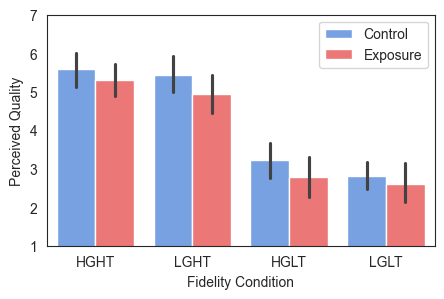}
    \caption{Barplot showing the perceived quality scores averaged across objects for all fidelity conditions in both groups. Error bars represent the 95\% confidence interval.}
    \label{fig:quality}
\end{figure}

\begin{table}
\centering
\caption{Results of the mixed-design ANOVA examining the effects of between-subjects and within-subjects on perceived quality. Significant effects are marked with an asterisk.}
\begin{tabular}{lcccc}
\textbf{Effect} & \textbf{df} & \textbf{F} & \textbf{$p$-value} & \textbf{Partial $\eta^2$} \\
\hline
\textbf{Between-Subjects Effects} & & & & \\
Exposure & 1 & 1.059 & 0.315 & 0.046 \\
\hline
\textbf{Within-Subjects Effects} & & & & \\
Object & 2 & 0.440 & 0.647 & 0.020 \\
Geometry & 1 & 4.623 & 0.043* & 0.174 \\
Texture & 1 & 100.549 & $<$0.001* & 0.820 \\
\end{tabular}
\label{tab:result_quality}
\end{table}

\subsection{Perceived Quality}
\paragraph{\textbf{Between-Subject Effect}}
No significant effect on perceived quality was observed between the exposure and control groups (see Table \ref{tab:result_quality}). On average, the exposure group scored lower across all conditions (M = 3.90, SD = 1.96) than the control group (M = 4.26, SD = 1.86) (see Figure \ref{fig:quality}). Robust analysis shows similar results, with Wald-Type Statistic (WTS) being 1.059 ($p$ = 0.303) and ANOVA-Type Statistic (ATS) also being 1.059 ($p$ = 0.306).

\paragraph{\textbf{Within-Subject Effects}} The type of object showed no significant effect on perceived quality (see Table \ref{tab:result_quality}). Both Geometry and Texture had significant effects, with Geometry exhibiting a large effect size and Texture an even larger effect size on quality perception (see Table \ref{tab:result_quality}). Specifically, the HGHT condition was rated best on average (M = 5.44, SD = 1.40), followed closely by the LGHT (M = 5.19, SD = 1.50). The HGLT was rated notably lower on average (M = 3.00, SD = 1.64), and the LGLT condition received the lowest score (M = 2.70, SD = 1.32) (see Figure \ref{fig:quality}). Robust analysis could confirm the findings for all three factors: Object (WTS = 0.834, $p$ = 0.659; ATS = 0.44, $p$ = 0.629), Geometry (WTS = 4.623, $p$ = 0.032; ATS = 4.623, $p$ = 0.032) and Texture (WTS = 100.549, $p$ < 0.001; ATS = 100.549, $p$ < 0.001).

\paragraph{\textbf{Interaction Effects}}
As no significant interaction effects between the factors were found for perceived quality, the results are excluded for conciseness.

\subsection{Perceived Realism}

\begin{figure}
    \centering
    \includegraphics[width=0.9\linewidth]{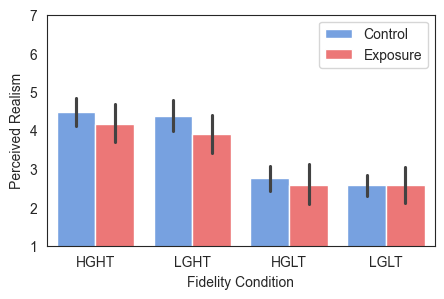}
    \caption{Barplot showing the perceived realism scores averaged across objects for all fidelity conditions in both groups. Error bars represent the 95\% confidence interval.}
    \label{fig:realism}
\end{figure}

\begin{table}
\centering
\caption{Results of the mixed-design ANOVA examining the effects of between-subjects and within-subjects on perceived realism. Significant effects are marked with an asterisk.}
\begin{tabular}{lcccc}
\textbf{Effect} & \textbf{df} & \textbf{F} & \textbf{$p$-value} & \textbf{Partial $\eta^2$} \\
\hline
\textbf{Between-Subjects Effects} & & & & \\
Exposure & 1 & 0.374 & 0.547 & 0.017 \\
\hline
\textbf{Within-Subjects Effects} & & & & \\
Object & 2 & 0.268 & 0.766 & 0.012 \\
Geometry & 1 & 1.592 & 0.220 & 0.067 \\
Texture & 1 & 53.705 & $<$0.001* & 0.709 \\
\end{tabular}
\label{tab:result_realism}
\end{table}

\paragraph{\textbf{Between-Subject Effect}}
No significant effect on perceived realism was observed between the prior exposure and control group (see Table \ref{tab:result_realism}). The control group received slightly higher scores on average (M = 3.54, SD = 1.40) compared to the exposure group (M = 3.30, SD = 1.69) (see Figure \ref{fig:realism}). Robust analysis confirmed the findings (WTS = 0.374, $p$ = 0.541; ATS = 0.374, $p$ = 0.543).

\paragraph{\textbf{Within-Subject Effects}}
No significant effect on perceived realism was found for the different Objects and for the changes in Geometry (see Table \ref{tab:result_realism}). However, Texture showed a significant effect with a very strong effect size \ref{tab:result_realism}. The HGHT condition received the best perceived realism scores (M = 4.32, SD = 1.34), with LGHT being only slightly lower (M = 4.14, SD = 1.45). The HGLT (M = 2.66, SD = 1.33) and LGLT (M = 2.57, SD = 1.20) conditions received low ratings (see Figure \ref{fig:realism}). Results from the robust analysis showed similar values: Object (WTS = 0.468, $p$ = 0.791; ATS = 0.268, $p$ = 0.759), Geometry (WTS = 1.592, $p$ = 0.207; ATS = 1.592, $p$ = 0.207) and Texture (WTS = 53.705, $p$ < 0.001; ATS = 53.705, $p$ < 0.001).

\paragraph{\textbf{Interaction Effects}}
As no significant interaction effects between the factors were found for perceived realism, the results are excluded for conciseness.

\subsection{Correlation Analysis}
Spearman rank tests revealed a strong positive correlation ($\rho = 0.802$, $p < 0.001$) between perceived quality and realism and a low positive correlation between age and perceived realism ($\rho = 0.231$, $p < 0.001$).

\section{Discussion}

\subsection{Effect of Prior Exposure}
The results indicate that $H_{1}$, concerning perceived quality, and $H_{2}$, concerning perceived realism, must both be rejected because no significant differences were found between the groups. While there is a tendency towards higher scores in the control group, the effect is weak and cannot necessarily be attributed to the absence of exposure to reference objects. Instead, the variation may have been caused by differences in prior experience with 3D virtual experiences, as the control group scored higher in all three categories (mixed reality, gaming, and 3D-modeling). The expectations of realism perception are likely based on past experiences \cite{yang2022}. Additionally, multiple reasons why the hypothesized effect did not occur can be considered. For example, participants may not have observed the physical objects sufficiently due to the implicit exposure, or upon entering VR, they may have become cognitively disconnected from the physical object. In fact, studies have shown that changing environments from physical to virtual negatively affects memory recall \cite{lamers2021}. Users may also have intentionally excluded them from their judgments and based their ratings purely on the capabilities and limitations of the VR headset.

\subsection{Effect of Fidelity Conditions}
Hypotheses $H_{3}$ and $H_{5}$ were supported by the data. Texture and geometry played a significant role in the perceived quality ratings of the DTs. As expected, higher resolution and higher levels of detail resulted in higher quality perception scores. Notably, the effect size of texture resolution (Partial $\eta^2$ = 0.820) was substantially more pronounced than that of geometry (Partial $\eta^2$ = 0.174). The reasons for this may be immediate visibility, i.e., the first impression. The reduced texture resolution resulted in an overall blurry appearance, whereas for reduced geometry, a closer observation was needed to recognize the missing details. It is also likely that the high-fidelity texture had a substituting effect on missing geometry, a well-known effect and practice for reducing mesh complexity \cite{rushmeier2000}.

In terms of perceived realism, only texture resolution resulted in a significant difference. Thus, hypothesis $H_{4}$ can be accepted, whereas $H_{6}$ must be rejected. The high-fidelity texture had only a slightly lower effect size (Partial $\eta^2$ = 0.709) on perceived realism than on perceived quality, whereas geometry had no significant effect. This suggests that participants perceived lower geometry as less qualitative, but they did not necessarily find the DT to be unrealistic. No prior research has explicitly measured the effect of texture resolution on perceived realism. However, if regarded as part of presence, the results align with the findings of heightened presence \cite{Hvass2017,Brade2021}.

These findings imply that in the development of DTs and virtual scenes for VR, texture resolution should be prioritized over geometric detail for optimization when aiming for a high degree of perceived quality and realism. An effective approach may include generating a high-fidelity photogrammetric model, followed by mesh simplification to reduce geometric complexity while preserving high-resolution textures.

\subsection{Correlation Findings}

The strong positive correlation found between quality and realism perception supports the known relationship between the two constructs, confirming $H_7$. With changing fidelity, perceived quality and realism increased and decreased together, with perceived quality reaching slightly higher scores. This aligns with models linking realism and quality perception \cite{JungLindeman2021,Fan2018}. However, it should be noted that these models focus on objective realism.

While the positive correlation between age and perceived realism was rather low, it aligns with findings from the literature \cite{Dilanchian2021}, confirming $H_8$. Following this argument, younger participants have higher expectations of 3D environments based on their familiarity with virtual worlds in digital media.

\subsection{Limitations and Future Work}
The non-significant effect of prior exposure may reflect insufficient statistical power due to the sample size of 24 participants. Given the small observed effect size, the study may have been underpowered to detect subtle between-subject differences, and a Type II error cannot be ruled out. Future studies with larger samples are necessary to determine whether exposure impacts quality and realism perception, as well as to mitigate the effects of group differences (e.g., gender balance, experience with 3D). 

Reflecting on the experimental design, we may have overestimated the impact of placing objects on the table. Instead, presenting the objects explicitly to the participants of the exposure group and considering giving them dedicated time for inspection before sending them to VR might yield clearer results. If a subsequent study using this explicit approach reveals an effect, the more subtle implicit setup could then be tested for comparison. Additionally, alternative forms of object knowledge could be considered beyond prior exposure, such as using widely recognizable objects.

More fidelity conditions should be tested to further deepen the understanding of optimal DT fidelity in VR experiences. Currently, only geometry and texture have been tested, neglecting the effects of material conditions. Moreover, the objects used were relatively simple in their geometry. In future studies, more complex models, different lighting conditions and material properties (e.g., shininess, shadows) should be tested.

Finally, follow-up experiments should increase the scale and complexity of DTs, shifting the focus from isolated objects to entire scenes or buildings. The aim is to assess whether the effects observed for fidelity conditions at the object level also apply at larger scales. Ultimately, the goal is to conduct studies with interactive, dynamic DTs with bi-directional data streams and continuously updated geometry. While the present study focused mainly on the visual quality of digital models, rating the quality of advanced DT properties is crucial for future research.

\section{Conclusion}
This study explored the effect of prior exposure on the quality and realism perception of DTs with varying texture resolution and geometric detail. The results showed no significant effect of prior exposure to physical objects on the perceived quality or realism of their DTs in VR. This could be due to cognitive disconnection upon entering VR, adaptation of user expectations to the VR, or the implicit and possibly insufficient exposure to the physical counterparts.
Instead, the fidelity conditions, particularly texture resolution, had a large effect on both quality and realism perception. High-resolution textures had a stronger impact than geometric detail, affecting users' immediate perception. Geometry influenced quality judgments but did not alter realism perception. These results indicate that to optimize DTs for VR usage, development efforts should prioritize high-fidelity textures over complex geometry. Mesh simplification paired with detailed textures could offer an efficient strategy to balance visual fidelity and computational performance.
While the exposure-related hypotheses could not be fully validated, the findings prompt critical questions about the perception of DTs in VR. They underscore the need to further investigate the cognitive link between physical objects and their virtual replicas, as well as how different fidelity levels influence this perception.

\bibliographystyle{ACM-Reference-Format}
\bibliography{main}

\end{document}